\documentclass[twocolumn,amsmath,amssymb,superscriptaddress,longbibliography,reprint]{revtex4-1}

\newcommand{\IMSS}{Muon Science Laboratory, Institute of Materials Structure Science, High Energy Accelerator Research Organization (KEK-IMSS), Tsukuba, Ibaraki 305-0801, Japan}
\newcommand{\IMSSKENS}{Neutron Science Division, Institute of Materials Structure Science, High Energy Accelerator Research Organization (IMSS, KEK), Oho, Tsukuba, Ibaraki 305-0801, Japan}
\newcommand{\Sokendai}{Graduate Institute for Advanced Studies, SOKENDAI}
\newcommand{\SRI}{Sumitomo Rubber Industries Ltd., Kobe, Hyogo 651-0072 Japan}
\newcommand{\STFC}{ISIS Facility, STFC Rutherford Appleton Laboratory, Chilton, Oxfordshire OX11 0QX, United Kingdom}

\usepackage{multirow}
\usepackage{graphicx}
\usepackage[dvipdfmx]{color}
\usepackage{dcolumn}
\usepackage[version=3]{mhchem}
\usepackage{txfonts}
\usepackage{bm}
\usepackage{ulem}

\usepackage[utf8]{inputenc}
\usepackage[T1]{fontenc}
\usepackage{mathptmx}
\usepackage{etoolbox}
\usepackage[hypertex,colorlinks=true,linkcolor=black,citecolor=blue,filecolor=blue,urlcolor=blue,setpagesize=false,nesting=true]{hyperref}

\makeatletter
\def\@email#1#2{%
 \endgroup
 \patchcmd{\titleblock@produce}
  {\frontmatter@RRAPformat}
  {\frontmatter@RRAPformat{\produce@RRAP{*#1\href{mailto:#2}{#2}}}\frontmatter@RRAPformat}
  {}{}
}%
\makeatother

\begin{document}
\title{Molecular dynamics of $cis$-polybutadiene across the glass transition\\ revealed by muonated-radical spin relaxation}

\author{S.~Takeshita}\affiliation{\IMSS}\affiliation{\Sokendai}
\author{H.~Okabe}\affiliation{\IMSS}
\author{M.~Hiraishi}\affiliation{\IMSS}
\author{K. M. Kojima}\thanks{Present address: Center for Molecular and Materials Science, TRIUMF, 
Vancouver, British Columbia, V6T2A3, Canada}\affiliation{\IMSS}
\author{A.~Koda}\affiliation{\IMSS}\affiliation{\Sokendai}
\author{H.~Seto}\thanks{Present address: Neutron Science and Technology Center, CROSS, Tokai, Ibaraki 319-1106, Japan}\affiliation{\IMSSKENS}
\author{T. Masui}\affiliation{\SRI}
\author{N.~Wakabayashi}\affiliation{\SRI}
\author{F. L. Pratt}\affiliation{\STFC}
\author{R.~Kadono}\thanks{ryosuke.kadono@kek.jp}\affiliation{\IMSS}
\date{\today}

\begin{abstract}
The local molecular motion of $cis$-polybutadiene, a typical polymeric material exhibiting a glass transition ($T_{\rm g}=168$ K), is investigated by the spin relaxation of muonated radicals, where the relaxation is induced by the fluctuation of hyperfine (HF) fields exerted from an unpaired electron to a nearby muon and surrounding protons. The relaxation rate $1/T_\mu$ measured under various longitudinal magnetic fields is analyzed using the recently developed theory of spin relaxation to consider the coexistence of quasistatic and fluctuating HF fields, where the fluctuation frequency for the latter $\nu$ is evaluated over a temperature $T$ range of 5--320 K. The obtained $\nu(T)$ is found to be well reproduced by the Arrhenius relation, and the activation energy and preexponential factor are in good agreement with those for the ``elemental process'' revealed by quasielastic neutron scattering and attributed to a fluctuation across three carbon-carbon bonds.  This result demonstrates that  muonated-radical spin relaxation is a promising approach for direct access to local molecular motions in the sub-nanosecond range and for their detailed modeling at the atomic scale.
\end{abstract}
\maketitle

\section{Introduction}
Glass transition is a phenomenon in which a fully or partially amorphous material changes from a flexible state exhibiting high viscosity and elasticity to a brittle state of glass-like or hard elasticity with decreasing temperature. Although the structure of the flexible state and the glassy state hardly changes above and below the transition temperature $T_{\rm g}$, its mechanical and thermodynamical properties often exhibit spectacular change (e.g., the viscosity changes by more than 10 orders of magnitude in a finite temperature range of a few tens of kelvins). It is extremely difficult to explain such huge changes in physical properties without accompanying structural changes, and the glass transition has been considered the last unsolved problem in condensed matter physics since the end of the last century \cite{Anderson:95}.

The glass transition is not only important as fundamental physics, but also significant in various application fields because it affects the physical properties, processability, and functionality of relevant materials. In particular, polymer materials are a storehouse of phenomena related to glass transition, and numerous studies have been conducted on those with a wide range of practical applications. As a result, it has been clarified that behind the glass transition, changes in molecular motion over an enormous range of spatial and temporal scales occur, which are now understood as relaxation phenomena rather than phase transitions \cite{Strobl:97,Kanaya:01,Donth:01,Ngai:03}.

Various experimental methods including quasielastic neutron scattering (QENS), nuclear magnetic resonance (NMR), dielectric relaxation, dynamic light scattering, and mechanical relaxation measurements have been used to identify several relaxation processes over a wide relaxation time: the ``boson peak,'' which is an excitation mode specific to amorphous materials, the ``fast'' process in the picosecond range (or the ``fast $\beta$'' process), the $\alpha$ process, and the Johari-Goldstein process (or the ``slow $\beta$'' process) which branches off from $\alpha$-process at low temperatures.  In addition, the ``elemental'' (E) process has been revealed in the subnanosecond range, which is regarded as a relaxation process specific to polymer glasses \cite{Kanaya:99}.  Among these relaxation processes, the microscopic understanding of the fast and E-processes is still in the early stage, partly due to the limited experimental techniques available for these time regimes. Nonetheless, these processes are important for applications, as they determine the fracture strength against mechanical impacts at lower temperatures \cite{Soles:21}.

Positive muons ($\mu^+$) implanted into a material behave as light radioactive isotopes of hydrogen (hereafter referred by the element symbol ``Mu'' as an isotope of hydrogen) \cite{Kadono:24a}. In particular, those implanted in an organic compound with unsaturated bonds will break the double bond by electronic excitation due to their kinetic energy (typically $\sim$4 MeV), forming a muonated radical state with an unpaired electron. Because of the hyperfine interactions between the unpaired electron and $\mu^+$, the muon spin relaxation ($\mu$SR) of muonated radicals is sensitive to spin fluctuations in the $10^8$--$10^{12}$ s$^{-1}$ region, which may serve as a useful probe for studying the above processes. Several pioneering $\mu$SR studies on polymers have shown that the kink observed in the temperature dependence of the spin relaxation rate occurs around $T_{\rm g}$, suggesting a link between them \cite{Pratt:00,Pratt:03,Pratt:05,Kanaya:15,Mashita:16,Pratt:16}. However, its microscopic origin has not always been clear.

In this paper, we show that the spin relaxation of Mu (muonated) radicals generated in $cis$-polybutadiene (PBD), a typical polymer glass, originates from the E-process. The fact that the activation energy for the fluctuation frequency is consistent with the potential barrier of intra-molecular rotation around carbon-carbon (CC) single bond and its correspondence to the results of molecular dynamics simulations provide strong evidence that the E-process is the conformation change in polymer chains as inferred from QENS \cite{Kanaya:99}. The mechanism of spin relaxation predicted from the local electronic structure of Mu radicals supports this interpretation. These results demonstrate that Mu-radical spin relaxation is a promising tool for studying molecular dynamics in the relevant time regime and for detailed modeling at the atomic scale.

 \section{$\mu$SR Experiment and DFT calculations}

\par
The PBD sample was prepared using material purchased from Ube Elastomer Co.~as received. The relative molecular weight of $cis$-, $trans$-, and $vinyl$-PBD were 97\%, 2\%, and 1\% (density = 0.91 g/cm$^3$). The number-average molecular weight $M_n$ was 1.9 MDa. The crystallization and glass transition temperatures were measured by specific heat measurements, yielding $T_{\rm cr}=254$ K and $T_{\rm g}=168$ K, respectively (the data are shown in the Supplemental Material (SM) \cite{SM}). The sample was shaped in a sheet with a dimension of $3\times2.5$ cm$^2$ $\times1.5$--2 mm, where the thickness was optimized for irradiation with a ``surface muon'' beam (an incident energy $E_\mu\approx4$ MeV, stopping range $\simeq0.1$ g/cm$^2$).

Conventional $\mu$SR measurements were performed on the PBD sample using the surface muon beam delivered to the Lampf spectrometer on the M20 beamline and HiTime spectrometer on the M15 beamline at TRIUMF, Canada. The $\mu$SR spectra [the time-dependent decay-positron asymmetry $A(t)$] which reflects the distribution of internal magnetic field (including hyperfine fields) at the $\mu^+$ site, was measured under a longitudinal field (LF; parallel to the initial Mu polarization ${\bm P}_\mu$) and weak transverse field (TF; perpendicular to ${\bm P}_\mu$) and were analyzed by least-squares curve fitting~\cite{musrfit}. Two sets of data were obtained by the Lampf spectrometer, one measured at 50, 190, and 300 K to investigate the detailed LF dependence of the $\mu$SR spectra and one at 5--320 K measured with the LF fixed at 20 mT to determine the temperature dependence of the spectra. The correction factor for the instrumental asymmetry under a given LF was calibrated by a separate set of LF-$\mu$SR measurements on a blank sample holder (made of silver), combined with the total asymmetry corresponding to 100\% muon spin polarization determined by a weak TF-$\mu$SR measurement. In addition, the spectra under a high transverse field (HTF $=2$ T) were measured with the HiTime spectrometer at 190--300 K to deduce hyperfine parameters of Mu radicals.

Calculations of the local molecular structure and hyperfine properties of the Mu radicals were performed on  a three unit oligomer in PBD using density functional theory (DFT) based on Gaussian16 \cite{Gaussian16} at the Becke three-parameter Lee-Yang-Parr (B3LYP)/correlation-consistent polarized valence double zeta (cc-pVDZ) level with molecular geometries determined using the PM7  semi-empirical method \cite{Stewart:13}. Quantum correction of the isotropic hyperfine coupling of the muon was done by calibration against the Mu radical in benzene.  Simulation of avoided level-crossing (ALC) resonance spectra and of LF-dependence for the initial asymmetry [$A(0)$ versus LF] at lower fields were done using the CalcALC program \cite{Pratt:22}.

\section{RESULT}

\begin{figure}[t]
  \centering
	\includegraphics[width=0.9\linewidth,clip]{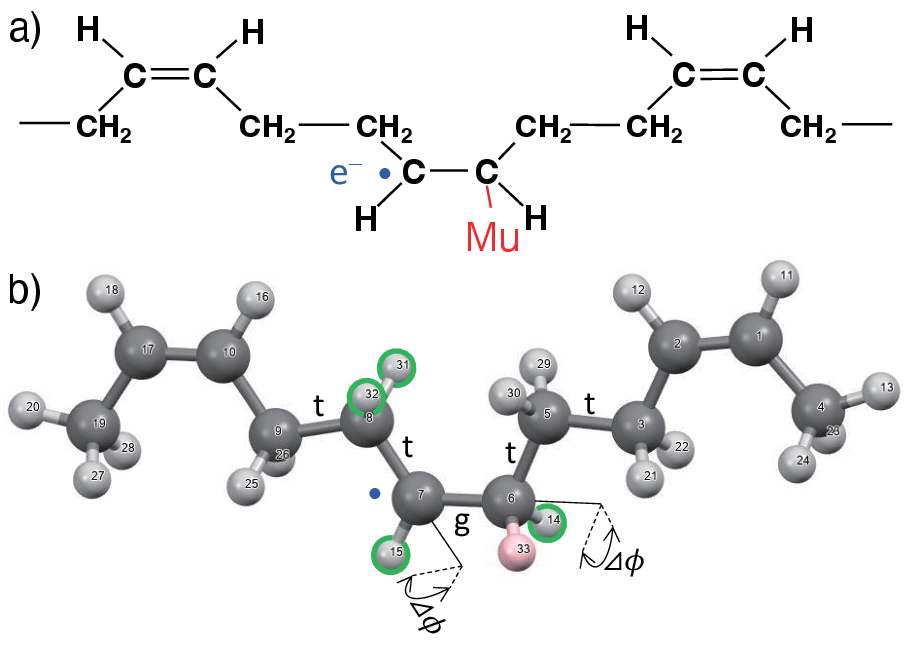}
	\caption{(a) The chemical formula for the muonated radical formed in PBD, and (b) the local structure obtained by DFT calculations for the three-unit oligomer. Mu (H33) is attached to C6 to induce an unpaired electron on C7 (blue dot), which undergoes HF interaction with Mu and NHF interaction with H14, H15, H31, and H32 (indicated by green circles).  ``t'' and ``g'' denote the CC bonds with trans and gauche structures, and $\varDelta\phi$ shows variation of the azimuth angle due to torsional/rotational motion around a CC bond. }
	\label{murad}
\end{figure}

As shown in Fig.~\ref{murad}(a), implanted $\mu^+$ cleaves the CC double bond to form the muonated radical. Its local electronic state is described by hyperfine (HF) interactions of an unpaired electron with $\mu^+$ and nuclear hyperfine (NHF) interactions with surrounding protons.  More specifically, the spin Hamiltonian for Mu radicals under an external magnetic field $B$ (parallel with $z$) is written as
\begin{equation}
\mathcal{H}/\hbar = 
\frac{1}{4}\omega_0{\bm \sigma}\cdot{\bf \tau} +\frac{1}{4}\omega_*({\bm \sigma}\cdot{\bm n})({\bm \tau}\cdot{\bm n}) -\frac{1}{2}\omega_\mu\sigma_z + \frac{1}{2}\omega_e\tau_z\nonumber 
\end{equation}\vspace{-6mm}
\begin{equation}
+\sum_m[\Omega_{\perp_m}{\bm S}_e\cdot{\bm I}_m +(\Omega_{\parallel_m}-\Omega_{\perp_m})S_e^zI_m^z]- \sum_m\omega_{n_m}I_m^z,\label{hn}
\end{equation}
where $\omega_0$ and $\omega_*$ respectively denote the transverse ($\omega_\perp$) and anisotropic ($\omega_\parallel-\omega_\perp$) parts of the HF interaction, $\omega_\mu=\gamma_\mu B$ with $\gamma_\mu=2\pi\times135.53$ MHz/T being the muon gyromagnetic ratio; $\omega_e=\gamma_e B$ with $\gamma_e=2\pi\times28024.21$ MHz/T being the electron gyromagnetic ratio; ${\bm \sigma}$ and ${\bm \tau}$ are the Pauli spin operators of the muon and electron; and ${\bm n}$ is a unit vector along the symmetry axis of the Mu-radical state \cite{Patterson:88,Yaouanc:10,MSR}. $\Omega_{\perp_m}$ and $\Omega_{\parallel_m}-\Omega_{\perp_m}$ are the transverse and anisotropic parts of the NHF parameter for the $m$th nuclear spin $I_m$ ($=\frac{1}{2}$ for $^1$H), and $\omega_{n_m}=\gamma_{n_m}B$ with $\gamma_{n_m}$ being the gyromagnetic ratio for the corresponding nuclei; the NHF parameters are usually dominated by the Fermi contact interaction, with an order of magnitude smaller contribution from the magnetic dipolar interaction (i.e., $\Omega_{\parallel_m}\gg\Omega_{\perp_m}$). 
While $\Omega_m$ are experimentally evaluated by the ALC resonance, LF-$\mu$SR spectra are determined by the linewidth \cite{Kadono:90},
\begin{equation}
\Delta_{\rm n}^2\approx\sum_m\frac{1}{3}(\Omega_{\parallel_m}^2+2\Omega_{\perp_m}^2)I_m(I_m+1),\label{Omg}
\end{equation}
where $m$ runs over the surrounding protons. 

The magnitude of the HF parameter obtained from HTF-$\mu$SR is $\omega_0/2\pi=314$--300 MHz at 250--300 K (which is weakly dependent on temperature and invisible below about $T_{\rm cr}$).  These values are consistent with earlier reports (311--300 MHz at 260--300 K) \cite{Pratt:00,Jestadt:99}, corresponding to 70\%--80\% of that evaluated by DFT calculations (374.7 MHz at 0 K).  The difference from the experimental values can be understood as being the result of thermally excited phonons (molecular vibrations). (For more details, see the SM \cite{SM}.)

\begin{figure*}[t]
  \centering
	\includegraphics[width=0.7\linewidth,clip]{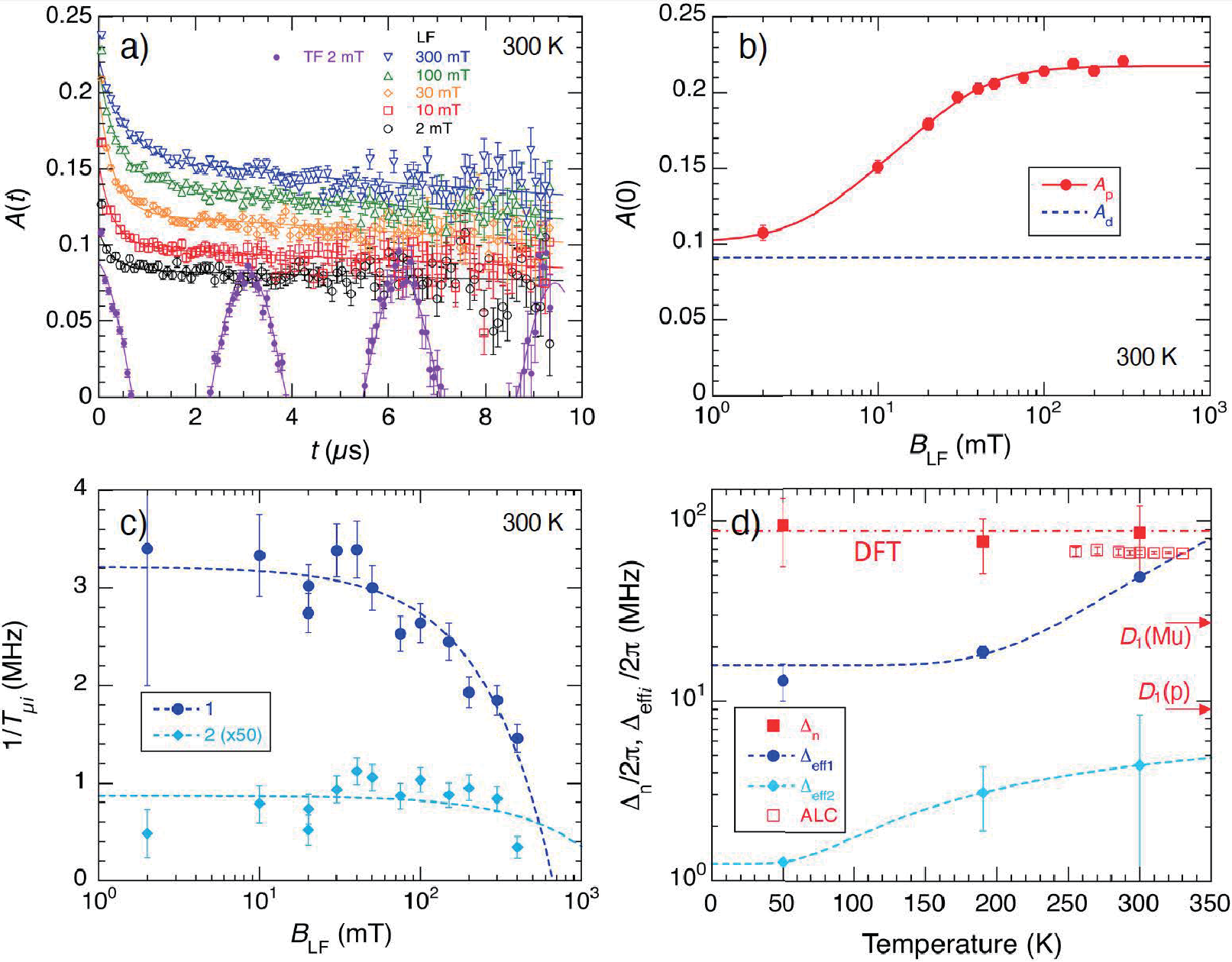}
	\caption{
	(a) TF- and LF-$\mu$SR time spectra observed at 300 K, which consist of paramagnetic ($A_{\rm p}$) and diamagnetic ($A_{\rm d}$) components. The solid curves represent the least-square fits by Eqs.~(\ref{Atf}) and (\ref{Alf}), from which the  LF dependence of (b) the initial asymmetry $A(0)$ and (c) spin relaxation rates $1/T_{\mu i}$ ($i=1,2$) are deduced. The horizontal dashed line in (b) shows $A_{\rm d}$, and dashed curves in (c) show the results of curve fits using Eq.~(\ref{tone}) (see text). (d) NHF parameter (linewidth) obtained from the detailed LF dependence of $\mu$SR spectra at 50, 190 and 300 K; $\Delta_{\rm n}$ from $A_{\rm p}$ vs $B_{\rm LF}$; $\Delta_{\rm eff1}$ and $\Delta_{\rm eff2}$ from $1/T_{\mu i}$ vs $B_{\rm LF}$; DFT calculations (dot-dashed line); and ALC results from Ref.~\cite{Jestadt:99}. Dashed curves for $\Delta_{{\rm eff}i}$ show fits for interpolation using the Arrhenius relation. Arrows labeled $D_1$(Mu) and $D_1$(p) indicate the dipolar terms in the HF/NHF parameters for Mu and protons predicted by DFT calculations, respectively.}
	\label{tspec}
\end{figure*}

Figure \ref{tspec}(a) shows typical $\mu$SR time spectra  observed at 300 K under several different LFs ($B_{\rm LF}=2$--300 mT) and a weak TF [$B_{\rm TF}=2$ mT; the portion for $A(t)>0$ is plotted]. These spectra consist of two components: one that shows the Larmor precession under $B_{\rm TF}$ and another component that exhibits gradual recovery of $A(0)$ with increasing $B_{\rm LF}$.  The former is represented by the equation
\begin{equation}
A(t)\simeq A_{\rm d}e^{-\lambda t}\cos\omega_\mu t,\label{Atf}
\end{equation}
which represents signals from the diamagnetic Mu states (Mu$^+$ or Mu$^-$ for isolated Mu) in the PBD sample and a background due to muons stopped in the materials around the sample [$=0.013(2)$]. $\mu$SR signals from Mu radicals under TF are quenched by fast depolarization due to the broad distribution of HF fields induced by the NHF interaction. The partial asymmetry of the Mu radical components in the LF-$\mu$SR spectra is determined to be $A_{\rm p}(B_{\rm LF})=A(0)-A_{\rm d}$, whose maximum value is $A_{\rm p0}\approx0.15$.

The lineshape of the LF spectra suggests that the Mu-radical signal consists of two components with different exponential relaxation rates.   In addition, a closer examination of the LF spectra near $t=0$ shows that $A_{\rm p}(B_{\rm LF})$ for $B_{\rm LF}\le10$ mT is much lower than $A_{\rm p0}/2$ expected for the spin-triplet radicals. This indicates that the unpaired electron is subjected to quasistatic NHF interactions with the surrounding protons. 
Based on these observations, the zero field (ZF)/LF spectra were analyzed by least-squares fitting with the following function \cite{Takeshita:24}:
\begin{equation}
A(t)\simeq \sum_{i=1,2}A_i\exp[-t/T_{\mu i}]+A_{\rm d},\label{Alf}
\end{equation}
where the partial asymmetry $A_i$ and the relaxation rate $1/T_{\rm \mu i}$ were free parameters. (The quality of the fits in Fig.~\ref{tspec}(a) can be confirmed by the residual plots and normalized  $\chi^2$ shown in the SM \cite{SM}.)
Considering the relatively large uncertainty in determining $A_1$ and $A_2$ independently, the total asymmetry $A_{\rm p}=A_1+A_2$ versus $B_{\rm LF}$ is analyzed using the following equations: 
\begin{eqnarray}
A_{\rm p}&=&A_{\rm p0}g_z(x_{\rm p}),\\
g_z(x_{\rm p})&=& \frac{\frac{1}{2}h_z(x_{\rm n})+x_{\rm p}^2}{1+x_{\rm p}^2},\:\:x_{\rm p}=2\gamma_{\rm av}B_{\rm LF}/\omega_0\label{gzp}\\
h_z(x_{\rm n})&\simeq& \frac{\frac{1}{3}+x_{\rm n}^2}{1+x_{\rm n}^2},\:\:x_{\rm n}\simeq\gamma_{\rm av}B_{\rm LF}/\Delta_{\rm n},
\end{eqnarray}
where $g_z(x_{\rm p})$ is the initial polarization of Mu radicals as a function of the normalized field $x_{\rm p}$ \cite{Patterson:88}, $h_z(x_{\rm n})$ is the approximated field dependence of the initial polarization under the NHF interaction characterized by the second moment $\Delta_{\rm n}$ (where the $\frac{1}{3}$ term corresponds to the possibility that the direction of the NHF field is parallel to the initial spin polarization of the triplet Mu$\dot{R}$ state \cite{Beck:75,Takeshita:24}), and $\gamma_{\rm av}=(\gamma_e+\gamma_\mu)/2= 2\pi\times14.08$ MHz/mT is the gyromagnetic ratio of the unpaired electron in the spin-triplet  state. 

The $B_{\rm LF}$ dependence of $1/T_{\mu i}$ was analyzed using the recently developed theory for the spin relaxation of paramagnetic Mu states that there is a possibility that only  part of the HF/NHF field is fluctuating, as was recently found in polythiophene (P3HT) \cite{Takeshita:24,Kadono:25}. Such a partial fluctuation of a local field ${\bm H}(t)$ is described by the autocorrelation function introduced by Edwards and Anderson for spin glasses, 
\begin{equation}
\frac{\langle {\bm H}(0){\bm H}(t)\rangle}{\Delta^2/\gamma_{\rm av}^2}\approx (1-Q_i)+Q_i\exp(-\nu_i t), \label{Acf}
\end{equation}   
where $\langle...\rangle$ denotes the thermal average over the canonical ensemble, $\Delta$ is the static linewidth given by $\Delta_{\rm n}$ and/or $\omega_0$ (and $\omega_*$), and
$Q_i$ is the fractional amplitude of the fluctuating component $(0< Q_i\le1)$ \cite{Edwards:75,Edwards:76}. 
 We assume that the spectral density of the spin fluctuation is described by that introduced by Bloembergen, Purcell, and Pound (BPP) in the field of NMR \cite{Bloembergen:48}, 
\begin{equation}
1/T_{\mu i}\approx\frac{\Delta_{{\rm eff}i}^2\nu_i}{\omega_{12}^2+\nu_i^2}\simeq\frac{\Delta_{{\rm eff}i}^2}{\omega_0}\left[\frac{\nu_i/\omega_0}{x_{\rm p}^2+\nu_i^2/\omega_0^2}\right],\:\:(i=1,2)\label{tone}
\end{equation}
where $\Delta_{{\rm eff}i}$ denotes the effective linewidth, $\nu_i$ is its fluctuation rate, and $\omega_{12}\simeq\gamma_{\rm av}B_{\rm LF}$ is the lowest intra-triplet transition frequency of Mu$\dot{R}$. The important point here is that $\Delta_{{\rm eff}i}^2$ is given by $\sqrt{Q_i}\Delta$ to reflect the partial fluctuation, whereas the $B_{\rm LF}$ dependence of $A_{\rm p}$ is dominated by the static ($1-Q_i$) component, yielding $\Delta_{\rm n}$ \cite{Kadono:25}.

The rough temperature dependence of $\Delta_{\rm n}$ and $\Delta_{{\rm eff}i}$ in Eqs.~(\ref{gzp})--(\ref{tone}) was evaluated with analyses of $A_{\rm p}$ and $1/T_{{\mu}i}$ versus $B_{\rm LF}$ for the spectra at 50, 190, and 300 K, for which detailed LF dependence was obtained.  As an example, the result for 300 K is shown in Figs.~\ref{tspec}(b)--\ref{tspec}(d).  The time spectra in Fig.~\ref{tspec}(a) are well reproduced by Eq.~(\ref{Alf}), yielding $A_{\rm p}$ and $1/T_{{\mu}i}$ at each $B_{\rm LF}$.  The LF dependence of $A_{\rm p}$ and $1/T_{{\mu}i}$ in Figs.~\ref{tspec}(b) and \ref{tspec}(c) is also in good agreement with Eqs.~(\ref{gzp})--(\ref{tone}). The deduced $\Delta_{\rm n}$ and $\Delta_{{\rm eff}i}$ obtained from these fits are shown in Fig.~\ref{tspec}(d). $\Delta_{\rm n}$ is least dependent on temperature, indicating that the radical state is stable over the entire temperature range of measurements. The mean value $\Delta_{\rm n}=2\pi\times84(18)$ MHz is consistent with previous ALC results of 66--68 MHz \cite{Pratt:00,Jestadt:99} within the range of error and that estimated from $\Omega_m$ from DFT calculations ($\simeq88$ MHz, which is dominated by the contribution of the four protons situated next to Mu; see the SM \cite{SM} for more details). Meanwhile, both $\Delta_{\rm eff1}$ and $\Delta_{\rm eff2}$ are significantly smaller than $\Delta_{\rm n}$ and dependent on temperature.

 \begin{figure}
  \centering
	\includegraphics[width=0.75\linewidth,clip]{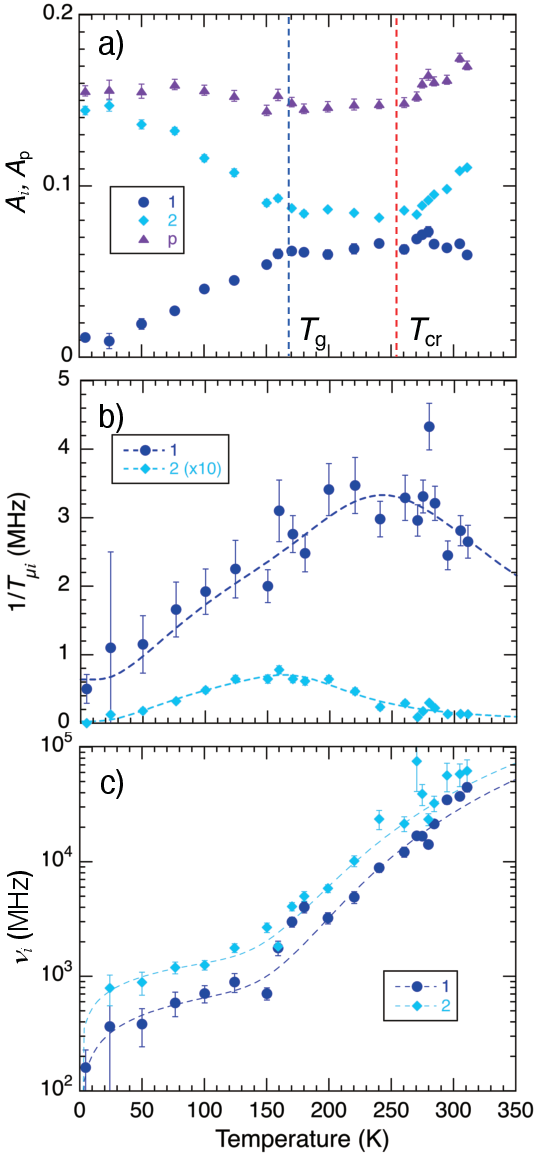}
	\caption{(a) Partial asymmetry $A_i$ and (b) relaxation rate $1/T_{\mu i}$ vs temperature deduced from curve fits of $\mu$SR spectra under LF $=20$ mT, where dashed curves in (b) show fits by the BPP model where $\nu_i$ is assumed to obey the Arrhenius relation (see text).  (c) Fluctuation frequency $\nu_i$ deduced from $1/T_{\mu i}$ using $\Delta_{{\rm eff}i}$ obtained by interpolation [shown in Fig.~\ref{tspec}(d)], where dashed curves show fits by Eq.~(\ref{Arrh}). }
\label{params}
\end{figure}

The detailed temperature dependence of these parameters was then obtained by analyzing the $\mu$SR spectra at ${\rm LF}=20$ mT at 5--320 K, and the result is shown in Figs.~\ref{params}(a) and \ref{params}(b).  Here, $A_1$ increases almost linearly with increasing temperature from 10 K and levels off just around $T_{\rm g}$. This suggests that the weight of the fluctuation component with large effective linewidth ($=\Delta_{\rm eff1}$) increases from low temperature to $T_{\rm g}$ and becomes approximately constant above $T_{\rm g}$. On the other hand, $A_2$ is in a trade-off relationship with $A_1$, decreasing to around $T_{\rm g}$ before becoming constant and then increasing again above $T_{\rm cr}$. These facts suggest that the Mu radicals probe the local molecular dynamics characteristic of the glass transition and crystallization of PBD.

 The temperature dependence of $1/T_{\mu i}$ has peaks around 250 K and 160 K, respectively, which, when viewed naively, appear to be also related to $T_{\rm cr}$ and $T_{\rm g}$. On the other hand, it is unlikely that the fluctuation frequency due to thermal excitation decreases at higher temperatures after reaching a maximum, and it is generally expected to keep increasing with increasing temperature. The dashed lines in Fig.~\ref{params}(b) show the result of fitting $1/T_{\mu i}$ versus temperature assuming the Arrhenius relation $\nu_i=\nu^*\exp(-E_a/k_BT)$ for $\nu_i$ in Eq.~(\ref{tone}) with varying $\Delta_{{\rm eff}i}$, which shows reasonable agreement with data. Thus it is strongly suggested that the peak of $1/T_{\mu2}$ corresponds to the $T_1$ minimum expected for $\omega_{12}=\nu_i$, and that the peak of $1/T_{\mu1}$ is due to the apparent shift of the $T_1$ minimum to a higher temperature due to the increase in $\Delta_{{\rm eff}1}$ for $T\gtrsim T_{\rm g}$ [as suggested in Fig.~\ref{tspec}(d)].
 
\begin{table}
\begin{tabular}{c|cccc}
\hline\hline
 & $\nu^*$ (s$^{-1}$) & $E_a$ (kcal/mol) & $c$ (s$^{-1}$K$^{-\kappa}$) & $\kappa$\\
 \hline
 $\nu_1$ & $3.1(9)\times10^{12}$ & 2.84(15) & $6.4(4.0)\times10^7$ & 0.50(14)\\
 $\nu_2$ & $2.8(9)\times10^{12}$ & 2.55(15) & $2.1(1.5)\times10^8$ & 0.39(16)\\
 \hline
$\Gamma$ (QENS) & $2.7\times10^{12}$ & 2.5 & -- & --\\
 \hline\hline
 \end{tabular}
 \caption{Preexponential factor and activation energy of the Arrhenius relation and parameters of the residual term in Eq.~(\ref{Arrh}) to reproduce the spin fluctuation rate of Mu radicals in PBD. The  Lorentzian linewidth $\Gamma$ in QENS spectra at $q=1.73$ \AA$^{-1}$ attributed to the E process and its activation energy for 295--425 K are quoted for comparison \cite{Kanaya:99}.}\label{tab1}
 \end{table}
 
However, it has become clear that attempting to deduce a large number of physical parameters with temperature variations directly from the fits of $1/T_{\mu i}$ vs temperature, as described above, is subject to large errors. Therefore, $\nu_i$ was derived using Eq.~(\ref{tone}) by fixing $\Delta_{{\rm eff}i}$ to the value interpolated for each temperature using the Arrhenius relation [dashed curves in Fig.~\ref{tspec}(d); see the SM for more details \cite{SM}], and the result is shown in Fig.~\ref{params}(c).
While $\nu_i$ increases monotonically at temperatures higher than $T_{\rm g}$, it tends to level off at lower temperatures. The dashed lines are the result of fits assuming the relationship 
\begin{equation}
\nu_i=\nu^*\exp(-E_a/k_BT)+cT^\kappa, \label{Arrh}
\end{equation}
where the last term is introduced to represent a contribution of residual phonon excitations: Assuming that Mu couples to the long-wavelength part of a Debye phonon spectrum, the fluctuation of HF/NHF fields via the spin-orbit interaction is expected to follow a power law in temperature \cite{Simanek:66,SM}. The obtained parameter values shown in Table \ref{tab1} indicate that both activation energy and pre-exponential coefficients are similar between $\nu_1$ and $\nu_2$.  The small $\kappa$ may be reasonable in the polymer, considering that it is proportional to the spatial dimension of the crystal lattice in the Debye model.

\section{DISCUSSION}\label{Dcn}

\begin{figure}[t]
  \centering
	\includegraphics[width=0.88\linewidth,clip]{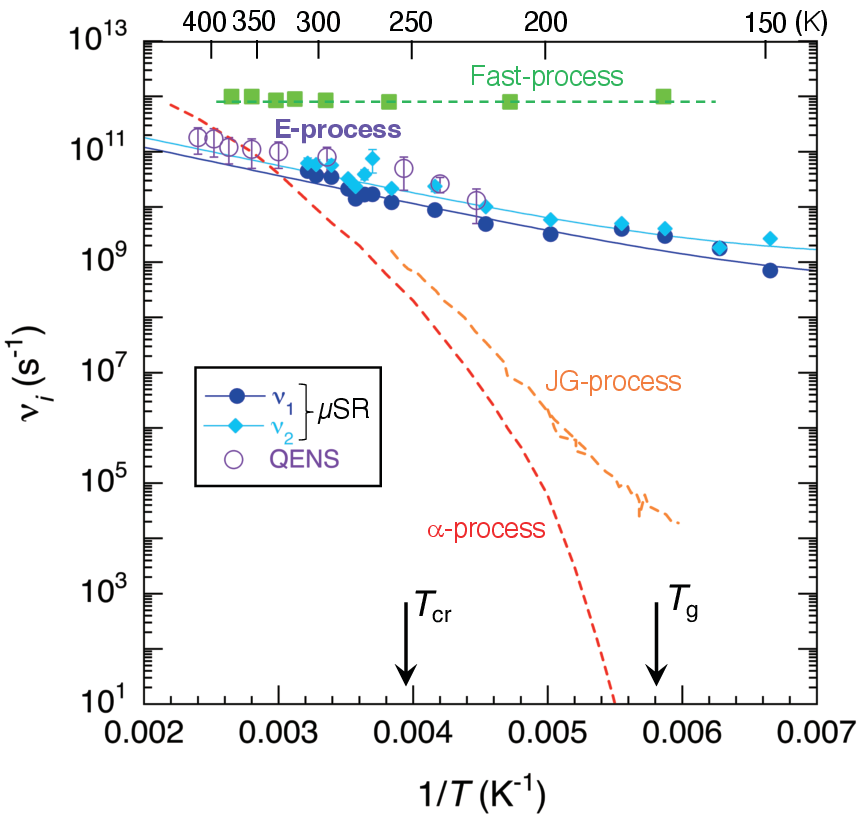}
	\caption{Fluctuation rate $\nu_i$ of Mu radicals plotted in the ``relaxation time map'' of PBD \cite{Kanaya:01}, with those observed using various experimental techniques are shown for comparison, for example, fast process and E process by QENS, Johari-Goldstein (JG) process by neutron spin echo, $\alpha$ process by viscosity, $^2$H-NMR, etc. }
\label{rlxmap}
\end{figure}

To explore the origin of the spin fluctuations observed by the Mu radical, we first plot the temperature dependence of $\nu_i$ in a relaxation time map showing the characteristic fluctuation frequencies of the various relaxation processes known so far versus inverse temperature in Fig~\ref{rlxmap}. As can be immediately inferred from Fig~\ref{rlxmap} and Table \ref{tab1}, the temperature dependence of both $\nu_1$ and $\nu_2$ is consistent with the E-process observed by QENS above $T_{\rm g}$. Thus, we can conclude that both signals from the radicals originate from the E-process in the corresponding temperature range.

A closer look at the QENS results shows that the scattering peak of the E process disappears below $T_{\rm g}$ \cite{Kanaya:99}. As seen in Fig.~\ref{params}, this coincides with the temperature region where $\nu_i$ begins to deviate from the Arrhenius relationship and is mainly dominated by the residual fluctuations described by the second term in Eq.~(\ref{Arrh}). Such behavior of $\nu_i$ suggests that the primary cause of fluctuations changes form the E-process to the long-wavelength phonons below $\sim$$T_{\rm g}$.  This provides another piece of evidence supporting the correspondence between QENS and Mu-radical spin relaxation.

The activation energies of the fluctuations obtained from QENS and the present results ($E_a$ in Table \ref{tab1}) are approximately the same as that associated with the CC single-bond rotation (e.g., 2.75 kcal/mol for ethane molecules). In addition, molecular dynamics (MD) simulations indicate that both single-bond rotations and counter-rotations of two adjacent bonds are possible in a polymer chain \cite{Gee:94,Kim:94,Okada:02}. This is because the distortion of the polymer chain caused by the single-bond rotation can be alleviated by the deformation degrees of freedom associated with the bonds adjacent to those involved in the conformation transition. In other words, the conformation transition of the polymer is localized by having such high deformation degrees of freedom. The MD simulations also revealed that the relaxation time of the conformation transition is $10^{-10}$--$10^{-11}$ s, and that the apparent activation energy of the conformation transition is almost identical to that of the CC single-bond rotation. These findings confirm that the E process is a conformation transition due to the single-bond motion on polymer chains.

The above microscopic model of fluctuations is also consistent with the local structure of the Mu radical predicted from the DFT calculations. As shown in Fig.~\ref{murad}(b), the C7 carbon (where the unpaired electron is located) is attached to both C6 and C8 via CC single bonds, which are further neighbored by C5 and C9 via single bonds as well. Such a local structure without double bonds has large deformation degrees of freedom, and the Mu radical signals can be interpreted as reflecting spin fluctuations associated with CC single bond rotations around the unpaired electron.

Furthermore, the origin of the observed two-component radical signals can be inferred to some extent from this structural model. The important factor here is the magnitude of $\Delta_{{\rm eff}1}$ and $\Delta_{{\rm eff}2}$, both of which are considerably smaller than $\omega_0$ and $\Delta_{\rm n}$ in the observed temperature range. This indicates that only part of the HF/NHF field is fluctuating, which may be understood by considering the fact that the dynamics of a PBD chain at low temperatures is dominated by torsional motion within the potential barrier of rotation; for a relatively small torsional angle $\varDelta\phi$,  $Q_i$ would be as small as $\langle\sin^2\varDelta\phi\rangle^{1/2}$ [see Fig.~\ref{murad}(b)].  Depending on their relative motion with the unpaired electron due to the way the trans and gauche bonds are arranged, Mu and protons around it will be divided into three groups, H31--H32, H15, and H14--Mu. Among these, H15 is expected to contribute little to the effective linewidth because it is attached to the same C7 site as the unpaired electron.  Moreover, a carbon atom put between trans and gauche bonds may be more mobile than that between two trans bonds. Considering these factors, it may be speculated that the $\Delta_{{\rm eff}1}$ and $\Delta_{{\rm eff}2}$ components correspond to the motion of C8 and C6, respectively.  As a more specific possibility for the cause of small $Q_i$, we suggest that $\Delta_{{\rm eff}i}$ may be dominated by the dipolar terms $\omega_*$ in the HF/NHF parameters, where $\Delta_{{\rm eff}1}$ and $\Delta_{{\rm eff}2}$ are respectively related to $D_1$ of Mu (28.3 MHz) and of protons ($\approx9$ MHz; see Fig.~\ref{tspec}(d) and the SM \cite{SM}).

Finally, we emphasize that such detailed modeling of the local motion in polymers is made feasible by the analysis of Mu radicals. In the preceding QENS studies, assuming that the Lorentzian  linewidth $\Gamma$ induced by the E-process in the scattering intensity $S({\bm q},\omega)$ follows a jump-diffusion model, the average jump distance is inferred to be $\langle\ell^2\rangle^{1/2}\simeq2.0$ \AA, which is attributed to a fluctuation across three CC bonds \cite{Kanaya:99,Kanaya:01}. The information from the Mu radicals not only is complementary to that from QENS but is also expected to allow a more detailed discussion of the local molecular dynamics as described above. In any case, in order to construct a more realistic model that explains the observed $\Delta_{{\rm eff}i}$, as well as to determine why the relative yields of the two relaxation components [$A_i$ in Fig.~\ref{params}(a)] correlate with $T_{\rm g}$ and $T_{\rm cr}$, further MD simulations of Mu-added PBD would be of great help.

\section{Summary and Outlook}
We showed that microscopic information on the local molecular motion of polymer chains can be extracted from the Mu-radical spin relaxation (MuSR) in PBD. 
It is now clear that the kink in the temperature dependence of the spin relaxation rate observed near $T_{\rm g}$ in previous $\mu$SR studies comes from the E process which is rapidly activated just above $T_{\rm g}$. This success may be partly due to the fact that PBD is one of the simplest polymers and that the electronic structure of the Mu radical is also relatively simple. Nonetheless, we already showed that MuSR can be applied to more complex organic systems, such as the side branched polymer P3HT  \cite{Takeshita:24} and the dielectric material, $tris$(4-acetylphenyl)amine (TAPA) \cite{Nakamura:25}. The window of fluctuation frequencies of $10^8$--$10^{12}$ s$^{-1}$ opened by this approach is a region previously covered only by QENS with particularly high energy resolution. It should also be stressed that  MuSR provides information complementary to QENS, in that it directly observes local dynamics in real space. Moreover, in many cases diamagnetic Mu states are observed in addition to radical states, as was the case in PBD. The spin relaxation of the latter provides information on the fluctuation frequency range of $10^4$--$10^9$ s$^{-1}$. Therefore, combining this with the information from MuSR, we can simultaneously obtain information on a wide time window of up to 8 orders of magnitude for the same specimen, which is expected to open up unprecedented possibilities for the study of glass transitions.

\begin{acknowledgments}
This work was supported by the Photon and Quantum Basic Research Coordinated Development (QBRCD) Program of the Ministry of Education, Culture, Sports, Science, and Technology (MEXT), Japan. The authors would like to thank T. Kanaya and H. Endo for helpful discussion. Thanks are also owed to the TRIUMF staff for the help during experiment conducted under Proposal No.~M1504. The DFT calculations were carried out on the STFC SCARF Compute Cluster.
\end{acknowledgments}

%
 
\end{document}